\documentclass[9pt,twocolumn,twoside]{osajnl}

\journal{ol} 

\setboolean{shortarticle}{true}

\usepackage{xcolor, soul} 
\definecolor{paleaqua}{rgb}{0.74, 0.83, 0.9}
\sethlcolor{paleaqua}

\title{Anisotropic Confinement of Chromophores Induces Second-Order Nonlinear Optics in a Nanoporous Photonic Metamaterial}
\author[1]{Karolina Waszkowska}
\author[1]{Pierre Josse}
\author[1]{Cl{\'{e}}ment Cabanetos}
\author[1]{Philippe Blanchard}
\author[1]{Bouchta Sahraoui}
\author[1]{Dominique Guichaoua}
\author[2]{Igor Syvorotka}
\author[3]{Olha Kityk}
\author[3]{Robert Wielgosz}
\author[4,5,6]{Patrick Huber}
\author[7,*]{Andriy V. Kityk}

\affil[1]{University of Angers, MOLTECH-Anjou, UMR CNRS 6200, 2 Bd. Lavoisier, 49045 Angers Cedex 01, France}
\affil[2]{SRC "Electron-Carat", 202 Stryjska str., 79031 Lviv, Ukraine}
\affil[3]{Energia Oze Sp. z o.o., ul. Częstochowska 7, 42-274 Konopiska, Poland}
\affil[4]{Technische Universität Hamburg, TUHH, Center for Integrated Multiscale Materials Systems CIMMS, Eißendorfer Str. 42, 21073 Hamburg, Germany}
\affil[5]{Deutsches Elektronen-Synchrotron, DESY, Center for X-ray and Nano Science CXNS, 22603 Hamburg, Germany}
\affil[6]{Universität Hamburg, Center for Hybrid Nanostructures CHyN, 22607 Hamburg, Germany} 
\affil[7]{Czestochowa University of Technology, Faculty of Electrical Engineering, Al. Armii Krajowej 17, 42-200 Czestochowa, Poland}

\affil[*]{Corresponding author: kityk@ap.univie.ac.at}

\doi{\url{http://dx.doi.org/10.1364/OL.416948}}

\begin{abstract}
Second-order nonlinear optics is the base for a large variety of devices aimed at the active manipulation of light. However, physical principles restrict its occurrence to non-centrosymmetric, anisotropic matter. This significantly limits the number of base materials exhibiting nonlinear optics. Here, we show that embedding chromophores in an array of conical channels 13 nm across in monolithic silica results in mesoscopic anisotropic matter and thus in a hybrid material showing second-harmonic generation (SHG). This non-linear optics is compared to the one achieved in corona-poled polymer films containing the identical chromophores. It originates in confinement-induced orientational order of the elongated guest molecules in the nanochannels. This leads to a non-centrosymmetric dipolar order and hence to a non-linear light-matter interaction on the sub-wavelength, single-pore scale. 
Our study demonstrates that the advent of large-scale, self-organised nanoporosity in monolithic solids along with confinement-controllable orientational order of chromophores at the single-pore scale provides a reliable and accessible tool to design materials with a nonlinear meta-optics. 
\end{abstract}

\setboolean{displaycopyright}{false}

\begin{document}

\maketitle


Second-order optical nonlinearity gives rise to three wave mixing processes including the linear electrooptic (Pockels) effect, second-harmonic generation (SHG), sum- or difference-frequency mixing as well as parametric generation and amplification \cite{Armstrong1962, Bloembergen1996}. These effects, observed first in inorganic crystal materials, have been demonstrated later also in a broad class of organic, organo-polymeric and organic-inorganic composite materials. In the last decades they have been in the focus of intensive experimental and theoretical studies, since they give rise to a large variety of novel nonlinear optical concepts and corresponding applications \cite{Zelikov}.  

 SHG consists in generating of optical waves of a doubled frequency 2$\omega$ induced by high intensity incident light of frequency $\omega$. By symmetry principles such a nonlinear optical effect, similarly as other three wave mixing processes, is allowed only in non-centrosymmetric media, e.g. polar and piezoelectric crystals. SHG can also be observed in doped polymer films that have previously been subjected to electrical poling, i.e. the exposure of a polymer film being heated to moderate temperatures to an high external electric field \cite{Sahraoui_2009}. This so-called ''corona poling'' induces a preferred orientation of the chromophores in thin surface layers of the polymer films inducing or enhancing thereby a second-order nonlinear polarization and thus nonlinear optical effects. 
 
Here, we first explore SHG induced by corona poling in a chromophore-doped polymer film and then we demonstrate that confinement of the identical chromophore in a nanoporous medium can also result in a composite exhibiting SHG - without corona poling. To that end we deposit the chromophore in parallel, tubular nanochannels 13~nm across in silica (\textit{p}SiO$_2$) membranes, see Fig.~1. The channels are of sub-wavelength size and thus allow us by modification of the channel filling to tune the mesoscale light-matter interaction and thus the resulting effective meta-optics \cite{Kadic2019, Sentker2018, Sentker2019, Kolmychek2020} of the nanoporous composite.  

The chromophoric dimer Ethane-1,2-diyl (2E,2'E)-bis(2-cyano-3-(5-(4-(diphenylamino)phenyl)thiophen-2-yl)acrylate) (hereafter termed D1, see structure in Fig.~1(a)) has been synthesized by linking two identical $\pi$-conjugated push-pull segments through a non-conjugated ethylene chain. The molecular structure of D1 has been confirmed by NMR spectroscopy and high-resolution mass spectrometry. 
HRMS (EI) calculated for C$_{54}$H$_{38}$N$_4$O$_4$S$_2$ 870.2334, found 870.2338 ($\Delta$= 0.4 ppm). UV-vis (CH$_2$Cl$_2$): $\lambda_{max}$ = 478 nm ($\varepsilon$ = 66500 L$\cdot$mol $^{-1}$$\cdot$cm$^{-1}$).

A strong electron-donating triphenylamine group (D) of the D1-dimer is linked to an electron-withdrawing cyanoacrylic ester group (A) by a thiophene ring as $\pi$-spacer. The D-$\pi$-A push-pull segments are characterized by an intramolecular charge transfer band observed in the visible spectral region. 

The red colored chromophore D1 was used as dopant (10 wt\%) for a poly(methyl methacrylate) (PMMA) host film deposited on a glass substrate, see Fig.~1(b).  5 mg of D1 and 50 mg of PMMA was dissolved in 1 ml of chloroform (CHCl$_3$). The solution was then deposited on an ultrasonically cleaned glass substrate of 1~mm thickness. Spin-coating deposition were performed by a spin coater (Spin200i, POLOS) at 2000~rpm. The thickness of the polymer doped film, characterized with a profilometer (Dektak 6M, Veeco) was 1020$\pm$20 nm. 

{The free standing \textit{p}SiO$_2$-membranes of 300 $\mu$m thickness were synthesized by thermal oxidation of electrochemically formed \textit{p}Si-membranes at $T=$800 $^o$C for 12~h under normal atmosphere \cite{Sentker2018}. The resulting membranes are characterized by an array of slightly conical nanochannels and an average pore diameter $2R=$13~nm, as determined by recording a volumetric N$_2$-sorption isotherms at $T =$ 77~K. The conicity of the pores results from the unidirectional \textit{p}Si etching process \mbox{\cite{Thelen2020}}. 

\begin{figure}[htbp]
\centering
\fbox{\includegraphics[width=0.86\linewidth]{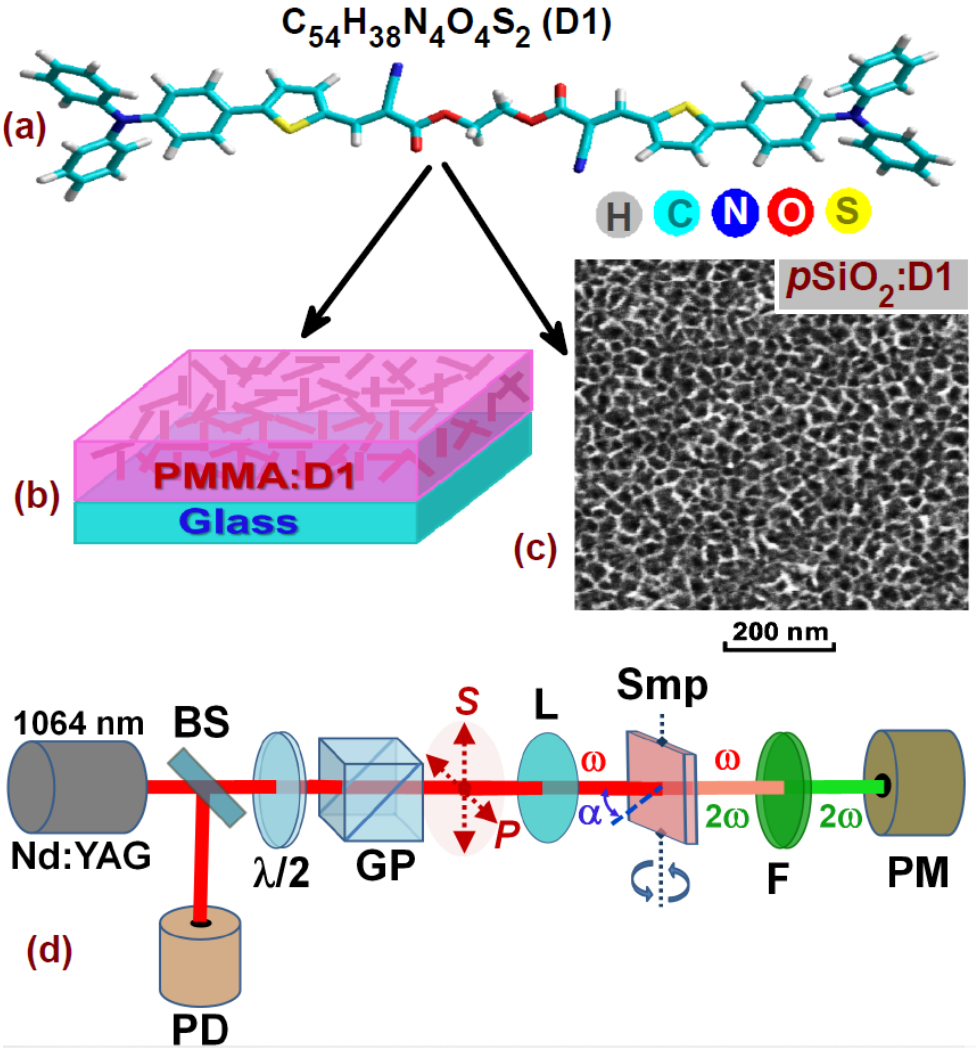}}
\caption{Molecular structure of the D1 chromophore (a) used as dopant for submicron polymer film (PMMA:D1) spincoated on the glass substrate (b) and as guest material for the host nanoporous silica membrane forming nanocomposite,  {\it{p}}SiO$_2$:D1 (c). Section (d) sketches the nonlinear optical setup (SHG experiment): laser Nd:YAG ($\lambda$=1064 nm, 100 $\mu$J, 30 ps,), halfwave plate $\lambda/2$, Glan polarizer GP, lense L, measured sample Smp, interference filter (532 nm) F, photodide PD, photomultiplier PM.}
\label{fig:fig1}
\end{figure}

\begin{figure}[htbp]
\centering
\fbox{\includegraphics[width=0.85\linewidth]{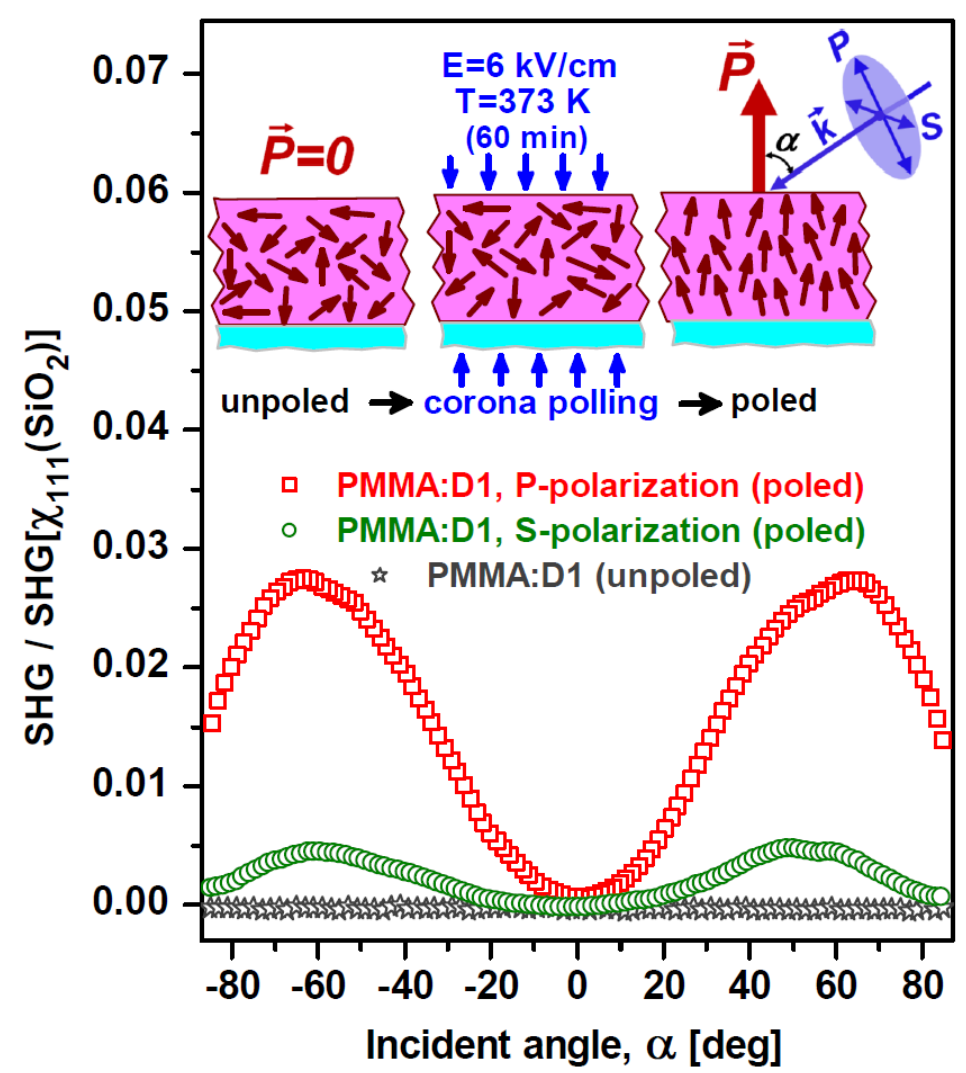}}
\caption{SHG response vs incident angle, $\alpha$, for $S$ and $P$ light polarizations measured in an poled and unpoled D1-doped submicron PMMA film, see labels. The insets sketch the corona poling procedure which leads to preferential  ordering of chromophore dipole moments parallel to the electric poling field resulting in appearing of macroscopic polarization $\vec{P}$.}
\label{fig:fig2}
\end{figure}

The guest chromophore dimer D1 has been deposited into the pSiO$_2$ by slow evaporation of its nearly saturated chloroform solution imbibed by the pore network, adding in each successive filling step equal portions of D1 chromophore ($\Delta f =$ 0.5\% of pore volume, $f$ is the fraction filling) as controlled and/or verified by optical absorption. 

In Fig.~1(d) the nonlinear optical setup used to record angular dependencies of the SHG response is sketched. The rotating $\lambda/2$-plate serves to set a polarization direction ($S$- or $P$-component) of the incident picosecond Nd:YAG-laser light ($\lambda$=1064 nm, 100 $\mu$J, 30 ps). The SHG-signal, selected by the interference filter F ($\lambda=532$ nm), is detected by a photomultiplier PM with a boxcar averager  (integration time 0.1 s) employing baseline subtraction. The measured SHG signal is normalized to the SHG response of a crystal quartz measured in $ooo$-geometry, i.e. $\chi_{111}($SiO$_2$) component.   

In Fig.~2 the SHG response for the chromophore doped polymer film is depicted. As expected, the unpoled PMMA:D1 film exibits no SHG effect as explainable by orientational disorder of the chromophores in the polymer host layer. The samples subjected to corona poling (external electric field of 6 kV/cm, time = 60 min, $T$=373 K) result in an SHG response appearing in both $S$- and $P$-polarization. The nonlinear optical conversion is anisotropic with a maximum SHG response at an incident angle $\alpha \approx$60$^o$, apparently dominating in $P$-polarization.

\begin{figure}[htbp]
\centering
\fbox{\includegraphics[width=0.99\linewidth]{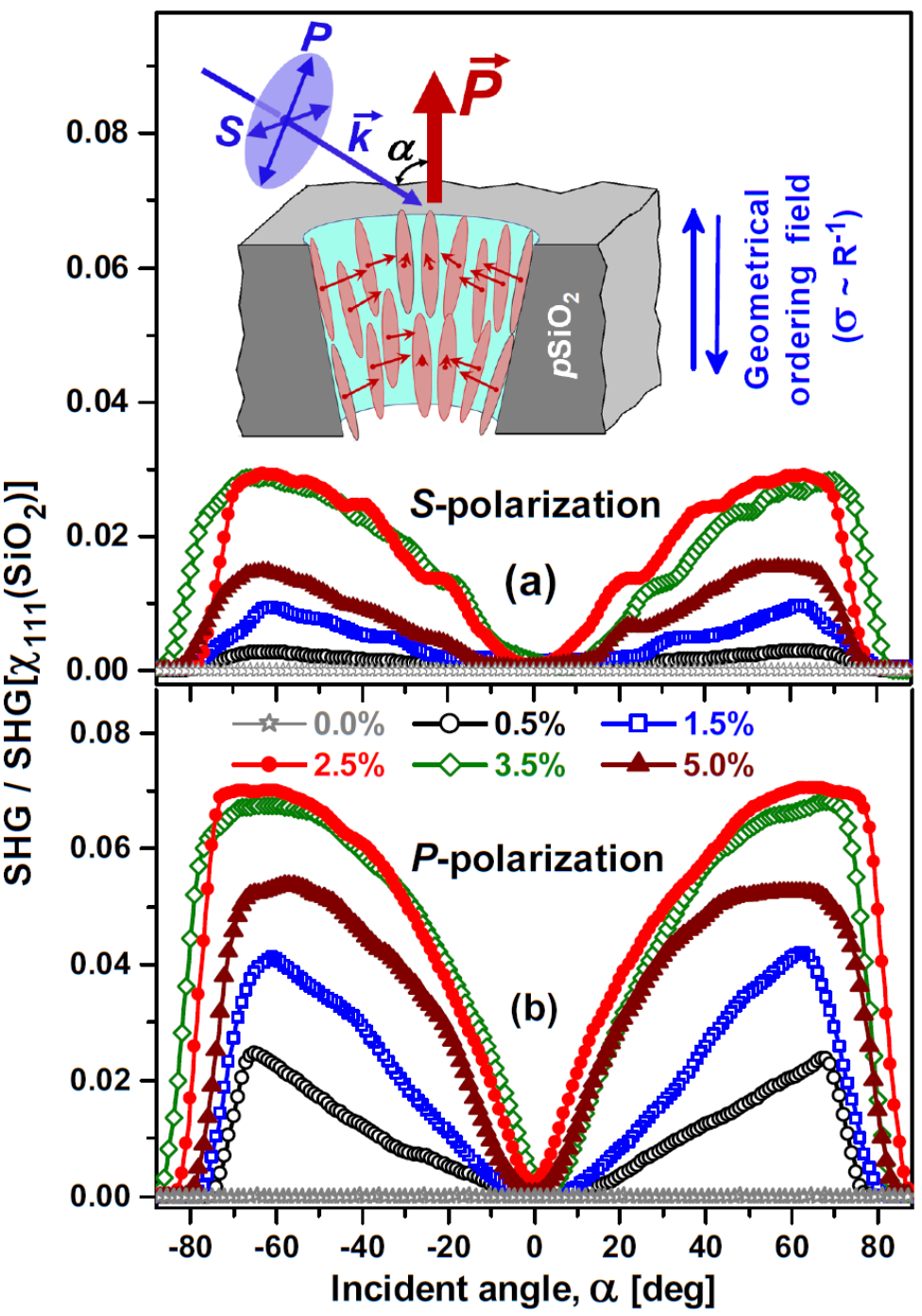}}
\caption{SHG response vs incident angle, $\alpha$, for $S$ (a) and $P$ (b) light polarizations measured in $p$SiO$_2$:D1 nanocomposite for different filling fractions $f$ of guest D1 chromophore, see labeled. Inset: Orientational ordering of elongated D1 molecules in conical nanochannels of mesoporous $p$SiO$_2$ membranes enforced by tubular confinement leads to a macroscopic polarization $\vec{P}$ parallel to channel axes.}
\label{fig:fig3}
\end{figure}

\begin{figure}[htbp]
\centering
\fbox{\includegraphics[width=0.85\linewidth]{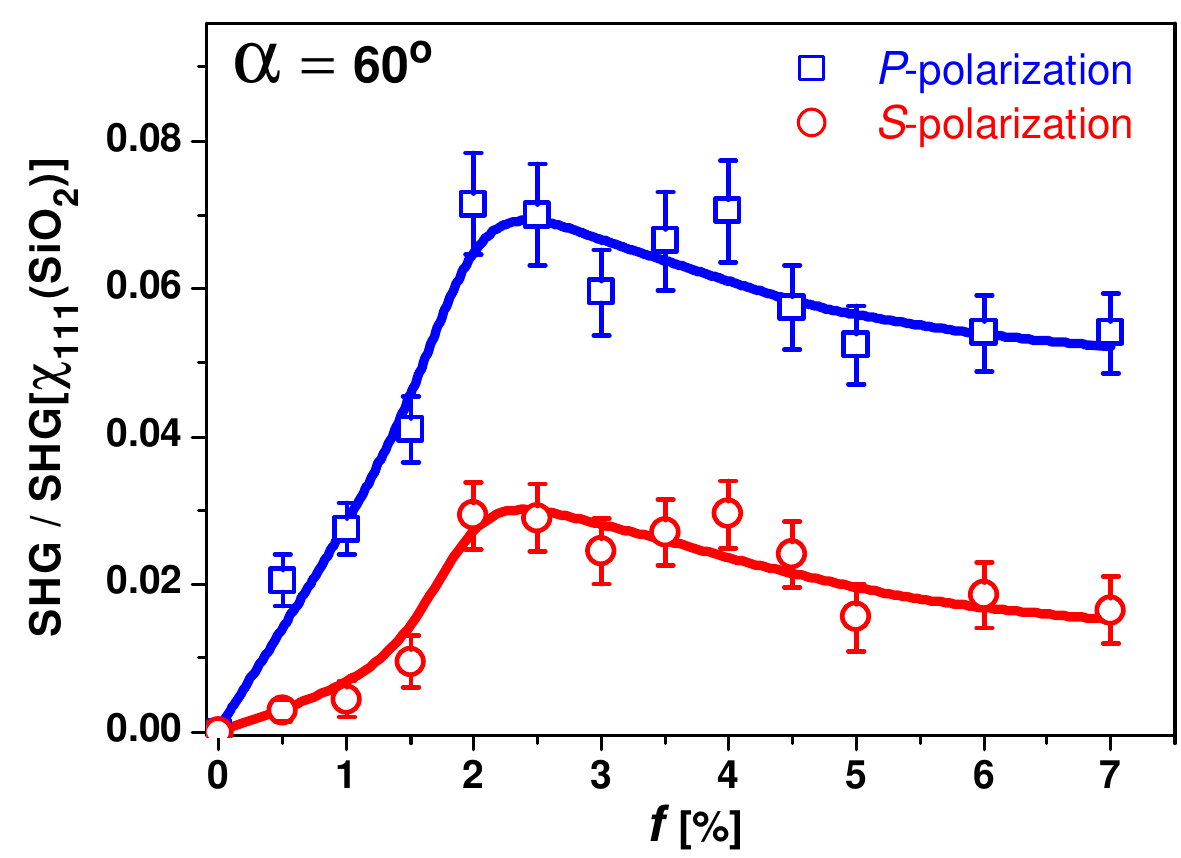}}
\caption{SHG response vs filling fraction, $f$, for $S$  and $P$ light polarizations measured in $p$SiO$_2$:D1 nanocomposite at the incident angle $\alpha$=60 deg.}
\label{fig:fig4}
\end{figure}

The $p$SiO$_2$-membrane  with chromophore coated pore walls, by contrast, exhibits the SHG response without poling, see in Fig.~3 the SHG response versus incident angle, $\alpha$, for $S$- (a) and $P$- (b) light polarizations measured from the  $p$SiO$_2$:D1 nanocomposite at different filling fractions of guest D1 molecules deposited in nanochannels. Similarly to the chromophore doped polymer film the SHG response of the $p$SiO$_2$:D1 nanocomposite exhibits a strong spatial anisotropy both with respect to the incident angle and the light polarization. The characteristics of these angular dependencies change somewhat upon increasing fractional filling $f$. However, in all cases both the measured $P$- and $S$-components of the SHG signal almost vanish at normal incidence ($\alpha$=0) and a maximum SHG response is detectable for incident angles, 50$<\alpha<$70$^o$. More surprising and unusual are the dependencies of the measured SHG signals on the fraction filling $f$. In Fig.~4 we present the SHG responses versus $f$ for $S$- and $P$-polarization components of the $p$SiO$_2$:D1 nanocomposite at an incident angle $\alpha$=60 deg as derived from the data presented in Fig.~3. At small $f$ both SHG signals rise reaching their maxima at $f\approx$2.5 wt\% and then gradually decrease with increasing $f$.

The measured optical absorption spectra of the $p$SiO$_2$:D1 hybrid, presented in Fig.~5, shed light on a potential reason for such a behavior. The characteristic feature of these spectra is a quite broad optical absorption band, centered at about 450~nm, strongly rising with concentration of deposited chromophore molecules in the mesoporous silica matrix, i.e. the fraction filling $f$ of the nanochannels. The narrow SHG emission at 532 nm appears in the region of the long-wavelength wing of this band, as marked by a vertical line in Fig.~5, and is considerably absorbed. In terms of the Lambert-Beer law the SHG intensity decays exponentially with the concentration of deposited molecules in nanochannels. Accordingly, at large filling fractions only a thin surface layer of chromophore doped silica matrix is expected to contribute effectively to the outcoming SHG light. Thus, one expects a saturation of the SHG intensity at larger $f$ and a subsequent decrease in agreement with the observations, see Fig.~5.

\begin{figure}[htbp]
\centering
\fbox{\includegraphics[width=0.77\linewidth]{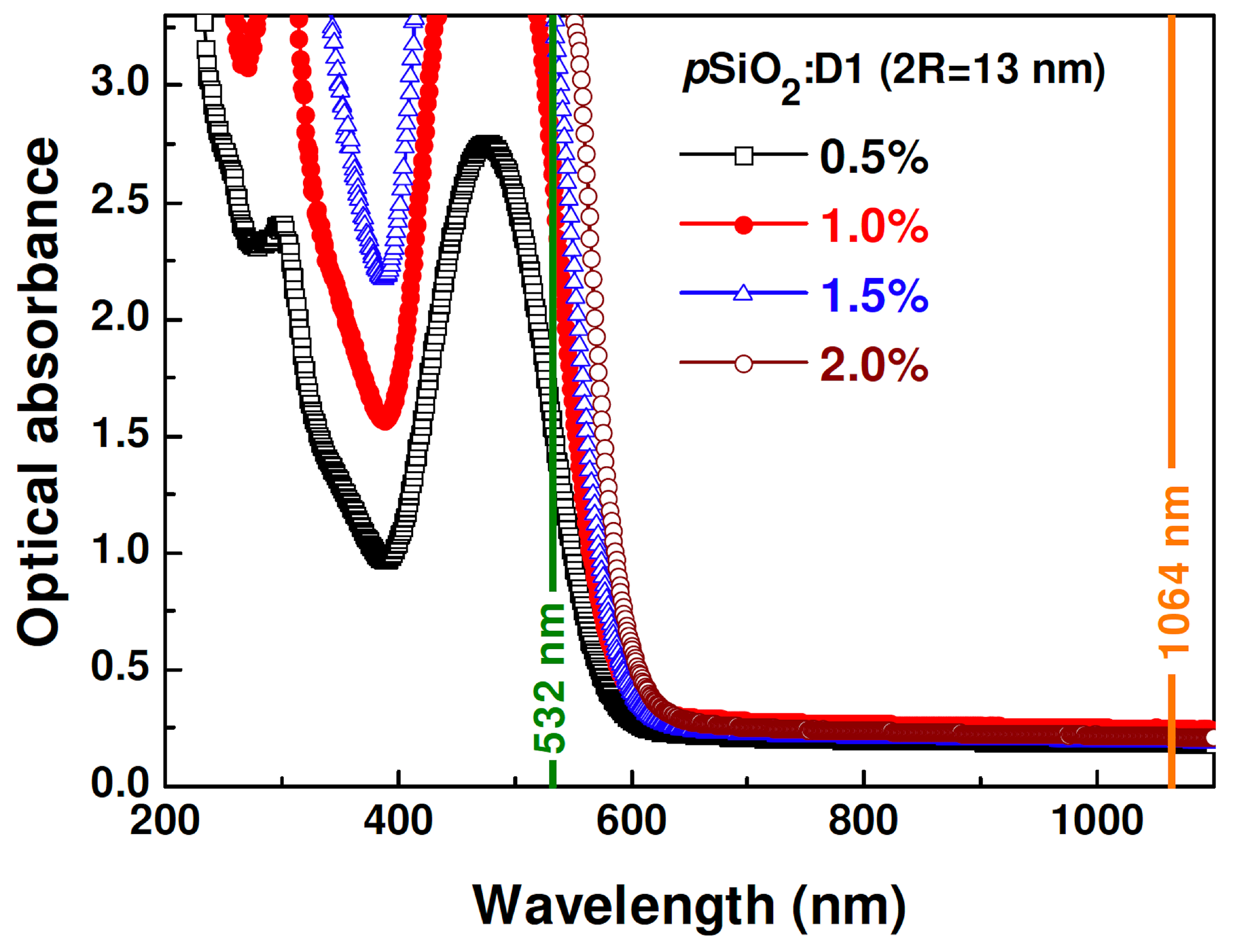}}
\caption{Optical absorbance vs filling fraction, $f$,  measured in $p$SiO$_2$:D1 nanocomposite at several filling fractions, see labeled. Vertical lines trace wavelengths of fundamental (1064 nm) and SHG radiations.}
\label{fig:fig5}
\end{figure}

The observation of an SHG effect in $p$SiO$_2$:D1 nanocomposite constitutes an interesting finding with regard to potential applications as nonlinear optical materials. It is however also a very interesting observation from a fundamental point of view and deserves a more detailed discussion. In the chromophore doped polymer film the second-order nonlinear polarization is induced  by corona poling, i.e. orientation and/or conformational distortion of chromophores molecules under a long term action of high electric field applied to the composite polymer film at temperatures close to the polymer glass transition point $T_g$. 

In the $p$SiO$_2$:D1 nanocomposite, however, the second-order optical nonlinearity likely follows from the geometrical confinement along with the conicity of the channels. From liquid-crystal based nanocomposites it is known that confinement in nanochannels effectively acts like a geometrical field \cite{Kityk1,Kutnjak1}. Particularly, it results for rod-like nematic molecules confined in cylindrical nanochannels to preferred molecular orientations along the channel axis in the pore wall proximity, i.e. the spatial constraints act similar as external electric or magnetic fields \cite{Kityk1,Kutnjak1}. Such a phenomenology may also explain the peculiar optical behavior of the elongated D1 molecules. Presumably, during the deposition they get preferentially oriented along the nanochannels, as sketched in the insert of Fig.~3(a). Note that it has been established for liquid crystalline systems that the strength of the confinement-induced molecular ordering field or capillary nematization \cite{Roij2000} is $\sigma \propto R^{-1}$. Thus, the effective moments enforcing axial alignment of the guest molecules are expected to be large here, since the average channel diameter ($2R=13$ nm) is only a few times the length of the chromophore molecules ($L_{D1}\sim$ 3.6 nm).

\begin{figure}[htbp]
\centering
\fbox{\includegraphics[width=0.65\linewidth]{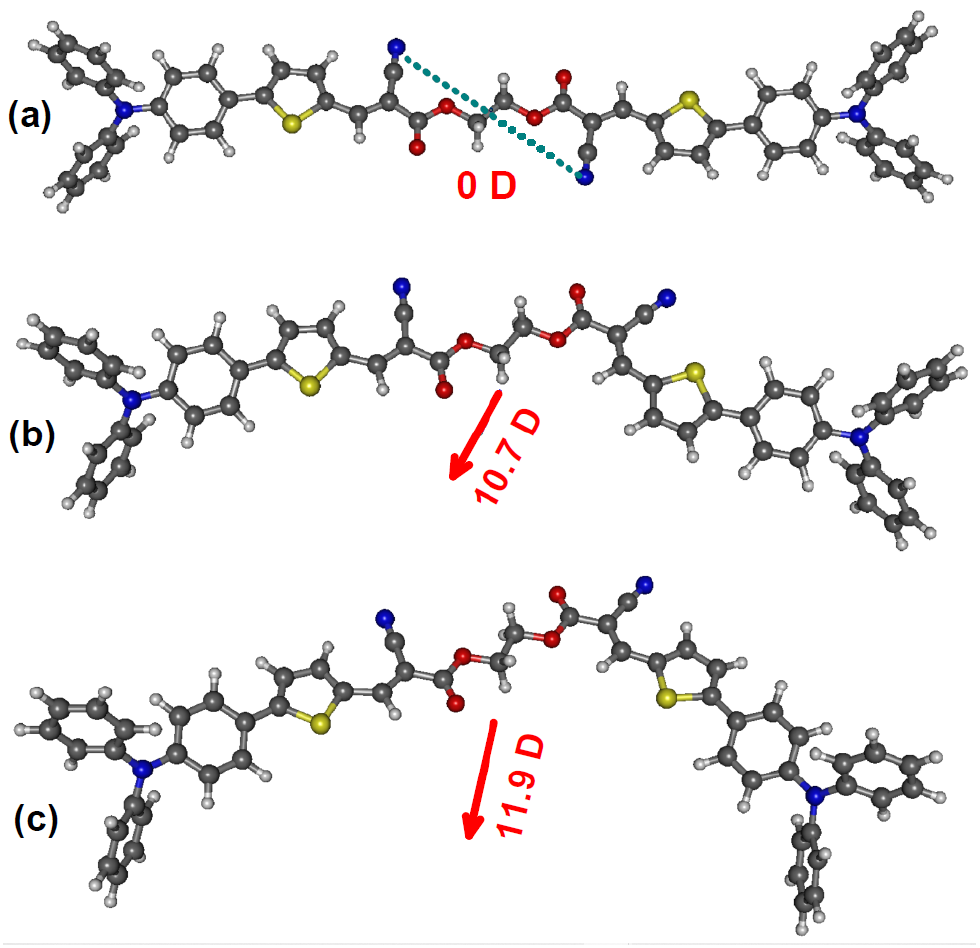}}
\caption{Examples of equilibrium (metastable) conformations of dimer molecule D1 obtained by DFT optimization in the ground state. (a) represents strait centrosymmetric conformation with zero permanent dipole moment ($\vec{P}=$0). (b) and  (c) are the two examples of bend shaped conformations characterized by large permanent dipole moment  ($\vec{P} \ne$ 0).}
\label{fig:fig6}
\end{figure}

A crucial issue in these considerations is the polar property of the guest molecules. For D1 it strongly depends on the conformation. In Fig.~6 we show examples of equilibrium conformations of the dimer molecule obtained by Density Functional Theory (DFT) optimization (CAM-B3LYP functional, 6-31+G(d,p) basis set) in the ground state using the Gaussian-16 quantum-chemical package \cite{Gaussian16}. Amazingly, the molecules characterized by the most stable centrosymmetric conformation (see Fig.~6(a)) have zero dipole moment ($\vec{P}=$0) and do not contribute to a second-order nonlinear polarization. Rotations of molecular segments about single bonds result, however, in a number of metastable non-centrosymmetric conformations. Some of them are characterized by quite large permanent dipole moments, as examplified in Figs.~6(b) and 6(c). Due to small energy barriers these metastable conformational states may be thermally activated or induced by interfacial interactions with the pore walls. Thus we suggest that this molecular polarity along with the confinement-induced alignment in conical channels results in a macroscopic polarization along the channel axis and thus causes the SHG effect, see Fig. \ref{fig:fig2}. A quantitative description of the SHG effect would necessitate a thorough  determination of confined, non-centrosymmetric D1 confirmations, which goes beyond of our dominantly experimental study.

In conclusion we demonstrate that it is possible to embed the chromophore dimer D1 into polymer PMMA films and mesoporous silica (pSiO$_2$) matrices with mean pore diameter of 13 nm. The synthesized PMMA:D1 and pSiO$_2$:D1 composites exhibit second-order nonlinear optical properties. The SHG response, absent in the unpoled chromophore doped polymer film, appears after long-term poling in high electric fields performed at temperatures close to the polymer glass transition point $T_g$. More importantly, the nanoporous pSiO$_2$ membranes with chromophore coated pore walls exhibit an SHG response without poling. We explain this by an orientational ordering of elongated guest chromophore molecules in the tubuluar, slightly conical nanochannels, i.e. as orientational alignment of chromophore molecules enforced by the geometrical confinement fields. 

From a more general perspective our study exemplifies how by filling of nanoporous media confinement-controlled ordering on the single-pore scale can be employed to tune the sub-wavelength light-matter interaction and thus the effective meta-optics of the hybrid materials. From a materials science perspective the resulting hybrids have an integrated optical functionality in combination with a mechanical rigidity provided by the nanoporous silica scaffold structure. Thus our study is a fine example, how self-organized nanoporosity in solids in combination with confinement-induced orientational order of organic chromophores results in materials with a non-linear meta-optics. 


\justify
\begin{Large} \sffamily{\textbf{Funding.}} \end{Large} {The presented results are part of a project that has received funding from the} European Union’s Horizon 2020 research and innovation programme under the Marie Sk{\l}odowska-Curie grant agreement No 778156. Support from resources for science in years 2018-2022 granted for the realization of  international co-financed project Nr W13/H2020/2018 (Dec. MNiSW 3871/H2020/2018/2). DFT calculations have been carried out using resources provided by Wroclaw Centre for Networking and Supercomputing (http://wcss.pl), grant No. 160. P.H. acknowledges support by the Deutsche Forschungsgemeinschaft within the SFB 986 ({\#}192346071).
\justify
\begin{Large} \sffamily{\textbf{Disclosures.}} \end{Large} The authors declare no conflicts of interest.



\end{document}